\documentclass[12pt,english]{article}
\usepackage{babel}
\usepackage{graphicx}

\begin{document}

\vspace*{-3cm}{\tt \noindent Contribution to the Seeheim Conference on
Magnetism (SCM 2001),  Seeheim, \\ Germany, Sept 9 -- 13, 2001, to 
appear in Physica Status Solidi A (2001)} 
\vspace{1.7cm}

\hspace*{1.5cm} \begin{minipage}{15cm}
\noindent{\bf\Large   Dipole   coupling   induced  magnetic   ordering
\\[0.2cm] in an  ensemble of nanostructured islands }  \\[0.5cm] P. J.
JENSEN$^{*,}$\footnote{Tel./Fax: +33 (0)561 55  68 33 / 60 65; e-mail:
jensen@irsamc.ups-tlse.fr}   and   G.   M.   PASTOR   \\[0.3cm]   {\it
Laboratoire de Physique Quantique,  Universite Paul Sabatier, UMR 5626
du CNRS \\ 118, route de Narbonne, F-31062 Toulouse, France} \\[0.5cm] 
\end{minipage} \vspace{0.7cm}

\noindent The magnetic ordering due  to the long range dipole coupling
in an ensemble of magnetic islands is investigated. If the islands are
large  enough and  closely separated,  the average  dipole  energy per
island can explain the  magnitude of the observed ordering temperature
of such an ensemble (U.\ Bovensiepen et al., J.\ Magn.\ Magn.\ Mater.\
{\bf192}, L386 (1999)).  The  energetical degeneracy with respect to a
continuous  in-plane rotation of  the magnetic  moments in  a periodic
ensemble of islands is lifted in presence of an island size dispersion
and an  irregular island  array. Many different  (metastable) magnetic
states are obtained, reminiscent  of a spin-glass behavior.  We obtain
that the average magnetic binding  energy per island due to the dipole
coupling  increases with increasing  positional disorder.   The island
ensembles  exhibit  non-collinear  magnetic structures,  resulting  in
non-saturated ensemble magnetizations.  The calculations are performed
with a  classical spin  model for ensembles  of islands in  unit cells
with  periodic  boundary  conditions.    The  point  dipole  sums  are
augmented by an island areal correction. \\

\noindent  {\bf  Introduction \  }  The  investigation of  interacting
magnetic nanoparticles is a very active field of current research both
experimentally   and  theoretically   \cite{Bov99}   --  \cite{Zha95}.
Measurements  on  growing  Co/Cu(001)  thin  films  exhibit  a  global
ordering  temperature of  the order  of 50  -- 100  K, and  a magnetic
hysteresis  and remanence  also  for coverages  below the  percolation
threshold \cite{Bov99}.  STM images taken in this coverage range yield
an ensemble of double-layered  Co islands with lateral size $\sim5$~nm
and   similar   island  separations,   which   should   behave  as   a
superparamagnetic  ensemble.   Blocking   effects  due  to  anisotropy
barriers estimated by the Arrhenius-N\'eel-model \cite{Nee49} could be
excluded  in this  temperature range.   In this  contribution  we will
examine  whether  the  long  range  magnetic dipole  coupling  can  be
responsible for a long range magnetic ordering at these temperatures.

It  is still  under debate  whether the  magnetic dipole  coupling can
induce   an   equilibrium    ordered   state   in   an   inhomogeneous
two-dimensional   (2D)   or   three-dimensional   (3D)   spin   system
\cite{And97}   --  \cite{Zha95}.   Certainly,   a  simple   ferro-  or
antiferromagnetic  state will not  evolve.  Instead  a spin-glass-type
\cite{Bin86} (random magnet)  is to be expected due  to the nonuniform
magnetic  interactions,  which  are  caused  by the  size-  and  shape
dispersion  as  well as  by  the  positional  disorder of  the  island
ensemble.  In  such a system  many different metastable  states exist,
which  are  often characterized  by  extremely  long relaxation  times
(magnetic viscosity).

The influence of the dipole coupling on the blocking effects have been
studied  extensively  in  the  Arrhenius-N\'eel-framework  for  single
particles, in particular for the investigation of dynamical processes.
The results are satisfactory for  weak interactions as compared to the
anisotropy barriers  of a single island  \cite{Cha96} -- \cite{Zal96}.
For strong interactions, however,  this model is no longer applicable.
Here we  will study the strongly interacting  case, considering solely
the dipole  interaction.  Note that  the long range  indirect exchange
(RKKY-)  interaction  might  also  be  responsible  for  the  observed
magnetic ordering in ensembles of magnetic islands.

Before describing the  model let us first estimate  whether the dipole
coupling  can  account for  the  observed  magnitude  of the  ordering
temperature $T_C$, which must be  of the order of the average magnetic
binding  energy per  island.   A pair  of  single spins  has a  dipole
interaction          strength           of          the          order
$E_{dip}\sim\mu_{at}^2/a_o^3\sim1$~K,  with $\mu_{at}\sim1\,\mu_B$ the
atomic   magnetic    moment,   $\mu_B$   the    Bohr   magneton,   and
$a_o\sim2.5$~\AA\  the   interatomic  distance  of   $3d$-  transition
metals. All $N$  spins of a magnetic island are  aligned by the strong
direct interatomic  exchange coupling, and  can be viewed as  a single
giant   spin   $M\sim   N\cdot\mu_{at}$   (Stoner-Wohlfarth   particle
\cite{StW48}).    The   diameter   of   a   flat   (2D)   islands   is
$L\sim\sqrt{N}$.    In   a   densely   packed  island   ensemble   the
island-island  separation  is $R  \sim  L$,  resulting  in an  average
magnetic    dipole    energy     per    island    pair    $E_{dip}\sim
M^2/R^3\sim(\mu_{at}^2/a_o^3)\;\sqrt{N}$.  Thus,  for a typical island
size   of   about   $N\sim1000$    atoms   the   binding   energy   is
$E_{dip}\sim30$~K, yielding  the correct  order of magnitude.   In the
following  we will  determine $E_{dip}$  for a  2D island  ensemble as
functions of the size dispersion and positional disorder. \\

\noindent {\bf Theory \ }  For the calculation of the magnetic binding
energy due to  the dipole coupling in a 2D island  array we consider a
unit  cell with  $n$  non-overlapping, disk-shaped  islands, and  with
periodic boundary conditions.  Within the unit cell the islands can be
placed  using  three different  types  of  arrangements: (i)  periodic
square array;  (ii) disturbed array,  the island centers  deviate from
the sites of  the periodic array using a  Gaussian distribution with a
standard deviation  $\sigma_r$; (iii) random setup.   The island size,
which determines its lateral extension, can be either the same for all
islands, of dispersed  around a mean value $\overline  N$ using also a
Gaussian distribution with a  standard deviation $\sigma_N$.  For such
an island ensemble we consider the following dipole-dipole Hamiltonian
with classical magnetic moments {\bf M}$_i$
($|{\bf M}_i|=N_i\cdot\mu_{at}$):
\begin{equation} 
{\cal  H}  =  \frac{1}{2}\;\sum_{i,j \atop  i\ne  j}\frac{1}{r_{ij}^5}
\left[{\bf  M}_i\;{\bf  M}_j\;r_{ij}^2-3\bigg({\bf r_{ij}}\,{\bf  M}_i
\bigg)\bigg({\bf r_{ij}}\,{\bf M}_j\bigg)\right] \;, \label{e1}
\end{equation}
$r_{ij}=|{\bf  r}_{ij}|=|{\bf  r}_i-{\bf  r}_j|$  being  the  distance
between the  centers of gravity of islands $i$ and  $j$.  The infinite
range of the  dipole interaction is taken into  account by applying an
Ewald type summation over all  periodically arranged unit cells of the
infinitely extended thin film. Since  for a large coverage the islands
are  closely separated,  we consider  in addition  to the  usual point
dipole   sum  the   so-called   areal  correction   (dipole-quadrupole
interaction).   This is  of the  order of  $(A_i+A_j)/r_{ij}^2$, where
$A_i\propto N_i$ is the area  of island $i$ \cite{PJJ}, and amounts up
to 50~\% of the average dipole energy.  For spherically shaped islands
this  correction vanishes  since  a sphere  has  no quadrupole  moment
\cite{YGP}.   We  neglect  here  the lattice  anisotropy.   The  shape
anisotropy due to the dipole  coupling within a single island vanishes
for disk-  or sphere-shaped islands.  For the  atomic magnetic moments
and  the interatomic distances  we choose  values appropriate  for Fe:
$\mu_{at}=2.2\;\mu_B$ and $a_o=2.5$~\AA.

Starting from an arbitrary initial  guess the total magnetic energy is
minimized  by varying  the magnetic  directions of  each  island. This
magnetic  relaxation  is  performed  with  the help  of  a  conjugated
gradient   method.    For   simplicity,   we   restrict   the   island
magnetizations    to    be    always    directed    in-plane,    ${\bf
M}_i=M_i(\cos\phi_i,\sin\phi_i,0)$,  characterized   by  its  in-plane
(azimutal) angle  $\phi_i$.  To account for the  many different energy
minima  in case  of  a  nonuniform spatial  island  arrangement or  in
presence of an island size dispersion, the magnetic energy is averaged
over many  different initial guesses for the  same island arrangement.
In addition,  we average $E_{dip}$ over ten  different realizations of
the  unit  cell,  using  the  same global  variables  (average  sizes,
standard deviations,  etc.)  characterizing the  island ensemble.  The
energy  reference  is  given  by  a  random  set  of  magnetic  angles
\{$\phi_i$\}, refering to a completely disordered system. \\

\noindent  {\bf Results  \ }  Before presenting  the  average magnetic
dipole  energy in  particular for  a disordered  island array,  let us
recall shortly the resulting structure for a uniform system, i.e.\ the
islands having  all the  same size  and placed on  a square  mesh. Two
different  magnetic  arrangements   are  possible:  (i)  a  metastable
parallel  magnetization of the  islands (ferromagnetic  solution), and
(ii) a columnar arrangement consisting of magnetized rows (or columns)
of islands  with alternating magnetic orientations. The  latter is the
ground state of a square array  of spins or magnetic islands, having a
vanishing remanent magnetization.  Note  that both solutions exhibit a
continuous  degeneracy with  respect  to an  in-plane rotation.   This
property is one  of the prerequisites for a  vanishing global magnetic
order   of    a   2D   magnetic   system    at   finite   temperatures
(Mermin-Wagner-theorem \cite{MeW66}).   By introducing now  a size- or
positional  disorder  into the  island  ensemble,  this degeneracy  is
immediately lifted. The system exhibits a (possibly very large) number
of discrete  states, which are  separated by energy barriers  and have
approximately  the  same energy  if  no  external  magnetic fields  or
aligned anisotropy easy axes are present.  The magnetic arrangement of
the island  ensemble is strongly non-collinear,  the net magnetization
is  small or  vanishes completely  (demagnetizing effect  due  to flux
closure),   cf.\  Fig.1.   We   emphasize  that   these  non-collinear
arrangements do not correspond to a disordered system, since the local
magnetizations are strongly correlated by the dipole interaction.  The
parallel and the columnar magnetic arrangements of the island ensemble
are always unfavorable, and even do no longer refer to a stable state.
Furthermore, due  to the  lifting of the  energy degeneracy  the above
mentioned  prerequisite  for the  Mermin-Wagner-theorem  is no  longer
fulfilled.  Therefore, a dipole  induced magnetic order might exist at
finite   temperatures   also    for   an   inhomogeneous   2D   system
('order-by-disorder-effect' \cite{Vil79}).

In  Fig.2 we  show the  average magnetic  dipole energy  $E_{dip}$ per
island as a function  of the island positional disorder, characterized
by  the standard deviation  $\sigma_r$. The  (average) island  size is
chosen to be $N=100$, the  average island-island distance is 1.5 times
the (average) island diameter, resulting in a coverage of about 35~\%.
$E_{dip}$ decreases with  increasing positional disorder, reaching its
minimal value  (= maximal  magnetic binding energy  per island)  for a
random setup of  the island ensemble.  The increase  of binding energy
with respect to  the uniform system amounts to roughly  50~\%.  As can
be also seen from Fig.2, an island-size dispersion has no large effect
on $E_{dip}$.

The reason for  the different effects of the two  types of disorder on
the magnetic binding  energy is that $E_{dip}$ is  a bilinear function
of the island sizes $N_i$, the size dispersion effect averages out for
symmetrically   distributed  island  sizes   around  the   mean  value
$\overline N$.  On  the other hand, the dipole  energy has a nonlinear
dependence    on   the    island-island    distance:   $E_{dip}\propto
r_{ij}^{-3}$.  Thus, with  increasing positional disorder the decrease
of $|E_{dip}|$ with an  enlarged distance $r_{ij}$ between some island
pairs  is more  than counterbalanced  by a  corresponding  increase of
$|E_{dip}|$  for smaller  distances between  other island  pairs. This
leads to  an increase  of the average  magnetic binding energy  of the
island ensemble. \\

\noindent {\bf Conclusion \ }  We have calculated the magnetic binding
energy  due to  the long  range dipole  coupling in  a 2D  ensemble of
magnetic islands.  The binding  energy increases by  ca.\ 50~\%  for a
random arrangement  of islands with  respect to a periodic  array.  An
island size dispersion  has less effect on the  binding energy. Due to
the nonuniform  system a large  number of different  metastable states
emerges   with  strongly   non-collinear  magnetic   arrangements,  as
reminiscent  of a  spin-glass  \cite{Bin86}.  We  expect  that if  the
islands are large enough and  closely separated, the dipole energy can
explain the  observed magnetic ordering  of a growing thin  film below
its percolation threshold \cite{Bov99}.  As a first approximation, the
resulting  ordering temperature  can be  determined from  a mean-field
approximation, and  corresponds to the spin-glass  temperature.  For a
comparison  with experiments  also  magnetic anisotropies  need to  be
taken into account, as well as overlapping magnetic islands \cite{PP}.
Furthermore,  one   can  calculate  the   ensemble  magnetization  and
susceptibility as a function of an applied magnetic field. \\

\noindent {\bf  Acknowledgement \ } The  authors acknowledge financial
support from CNRS (France), and  by the EU GROWTH project AMMARE under
contract number G5RD-CT-2001-00478.

\begin{figure} 
\hspace*{1cm}  \includegraphics[bb=90   330  530  790,clip,width=15cm]
{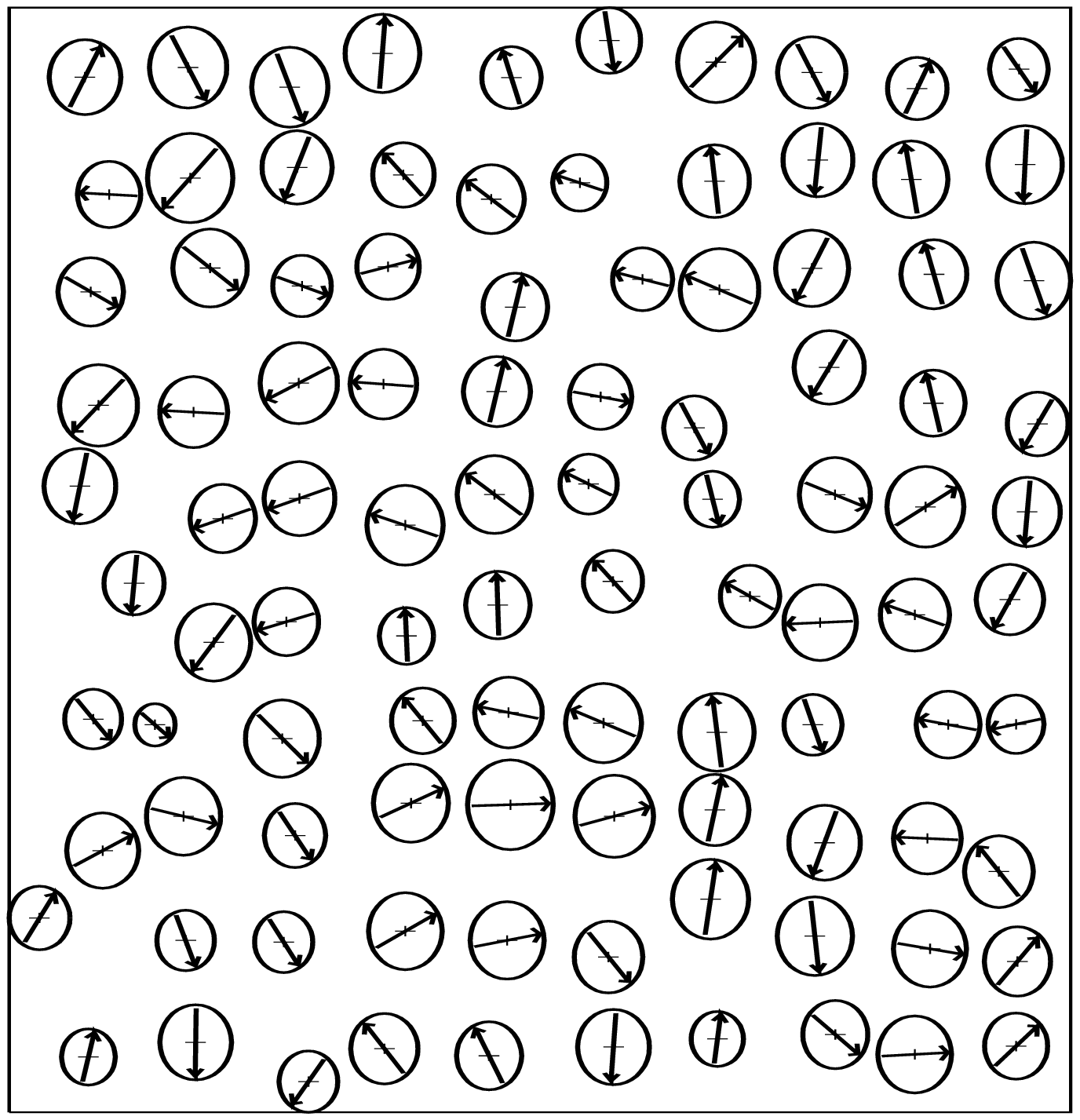}

\caption{Visualization   of   a   typical  (meta-)   stable   magnetic
arrangement in the unit cell as obtained from the calculations for the
average dipole  energy.  The average island size  is $\overline N=100$
spins, yielding an average island radius of about $10\;a_o$, $a_o$ the
interatomic  distance.  The  size of  the unit  cell is  chosen  to be
$(150\;a_o)\times(150\;a_o)$, containing  100 islands, and  yielding a
coverage   $\Theta\sim35$~\%.     The   spatial   island   arrangement
corresponds to a disturbed array,  the island centers deviate from the
sites of  a square array  with a standard  deviation $\sigma_r=10$~\%.
The island size exhibits  a dispersion of about $\sigma_N=20$~\%.  The
arrows depict  the in-plane  directions of the  island magnetizations.
Note the tendency  to the formation of island  chains and flux closure
structures.}
\end{figure} \vspace{1cm} 

\begin{figure} 
\hspace*{0.5cm} \includegraphics[width=13cm,angle=-90]{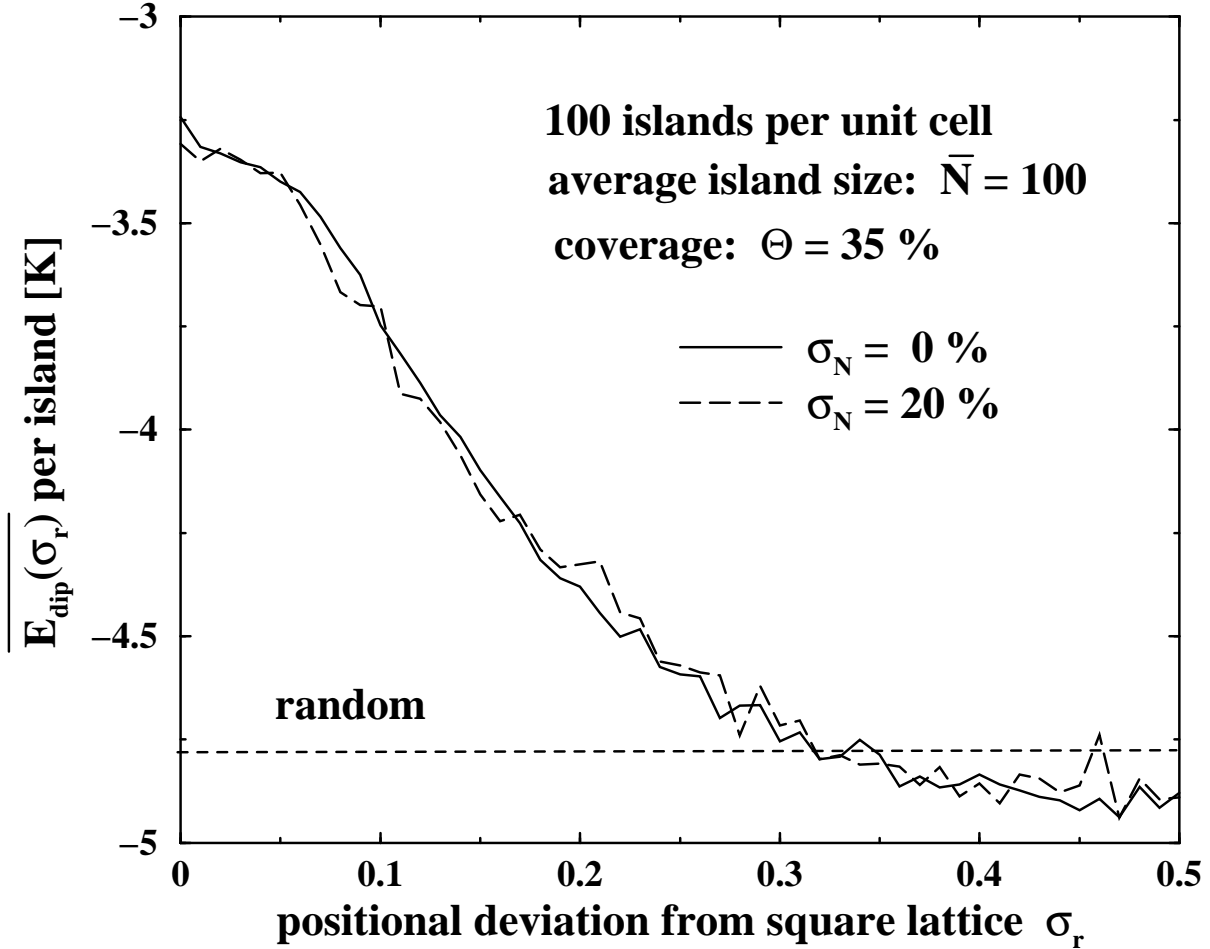}

\caption{Average  dipole   energy  $\overline{E_{dip}(\sigma_r)}$  per
island  of a  2D ensemble  of magnetic  islands as  a function  of the
positional  disorder,  which is  measured  by  its standard  deviation
$\sigma_r$ from a  square array, s.\ Fig.1. The full  line refers to a
system with the same size for all islands, the dashed line to a system
with 20~\% size  dispersion. The  line denoted  by 'random'  refers to
$\overline{E_{dip}}$ of a random  island arrangement in the unit cell.
For the  atomic magnetic moment  and the interatomic distance  we have
choosen   values  appropriate   for   Fe:  $\mu_{at}=2.2\;\mu_B$   and
$a_o=2.5$~\AA. }
\end{figure}
 
\end{document}